\documentclass[%
 reprint,
superscriptaddress,
 amsmath,amssymb,
 aps,
 prl,
]{revtex4-2}

\usepackage{amsmath}
\usepackage{array}
\usepackage{mathtools}
\usepackage{siunitx}
\usepackage{physics}
\usepackage{braket}
\usepackage{bbold}
\usepackage{tikz}
\usepackage{graphicx}
\usepackage{dcolumn}
\usepackage{bm}
\usepackage{placeins}
\usepackage{xcolor}
\usepackage{bibentry}
\usepackage{ulem}
\usepackage[colorlinks,linkcolor=black]{hyperref}
\usepackage{orcidlink}

\newif\ifshownotes
\shownotestrue

\makeatletter
\def\Ddots{\mathinner{\mkern1mu\raise\p@
\vbox{\kern7\p@\hbox{.}}\mkern2mu
\raise4\p@\hbox{.}\mkern2mu\raise7\p@\hbox{.}\mkern1mu}}
\makeatother
\def \FigureOne
 {
\begin{figure*}[t]
 \centering
  \includegraphics[width=\linewidth]{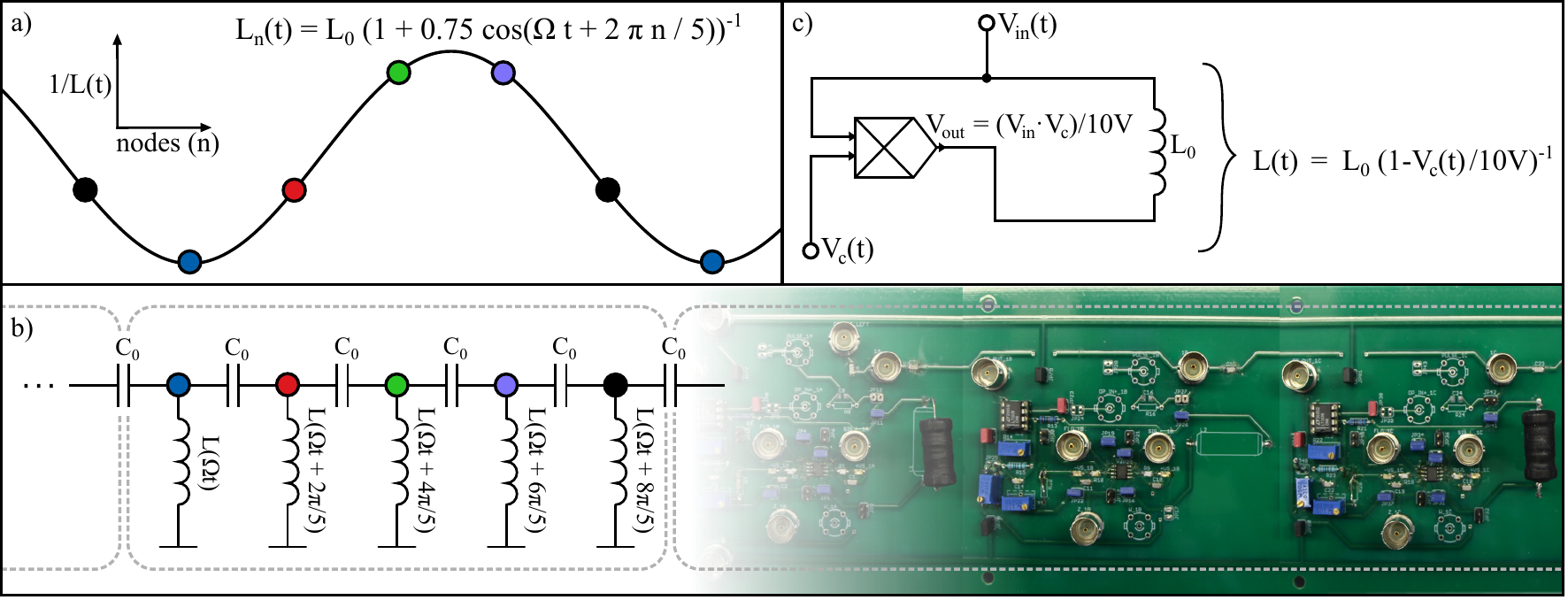}
 \caption{Implementation of the Thouless Pump in a Circuit. 
 a) Variation of the inductance over space (nodes) and time. The
 inductors act as a cosine-shaped potential shifted over time to
 induce adiabatic pumping. b) The circuit consists of
 a chain of capacitors, with each node connected to ground by a
 variable inductor. The dashed box indicates the unit cell of a
 periodic chain. On the left, a simplified circuit diagram of one unit cell is depicted, which blends into a photograph of a circuit board on the right, showing three circuit nodes. A detailed description of the circuit board can be found in the supplement. The complete circuit is made up of eight unit cells (40 nodes).  c) Schematic of the time-variable inductor. An analog
 multiplier creates a signal applied to an inductor connected to the
 node. The current flowing in and out of the sub-circuit thus mimics that of a time-variable inductor to ground.  
 }
 \label{Fig:Fig1}
\end{figure*}
}

\def \FigureTwo
 {
\begin{figure*}[t]
 \centering
 \includegraphics[width=\linewidth]{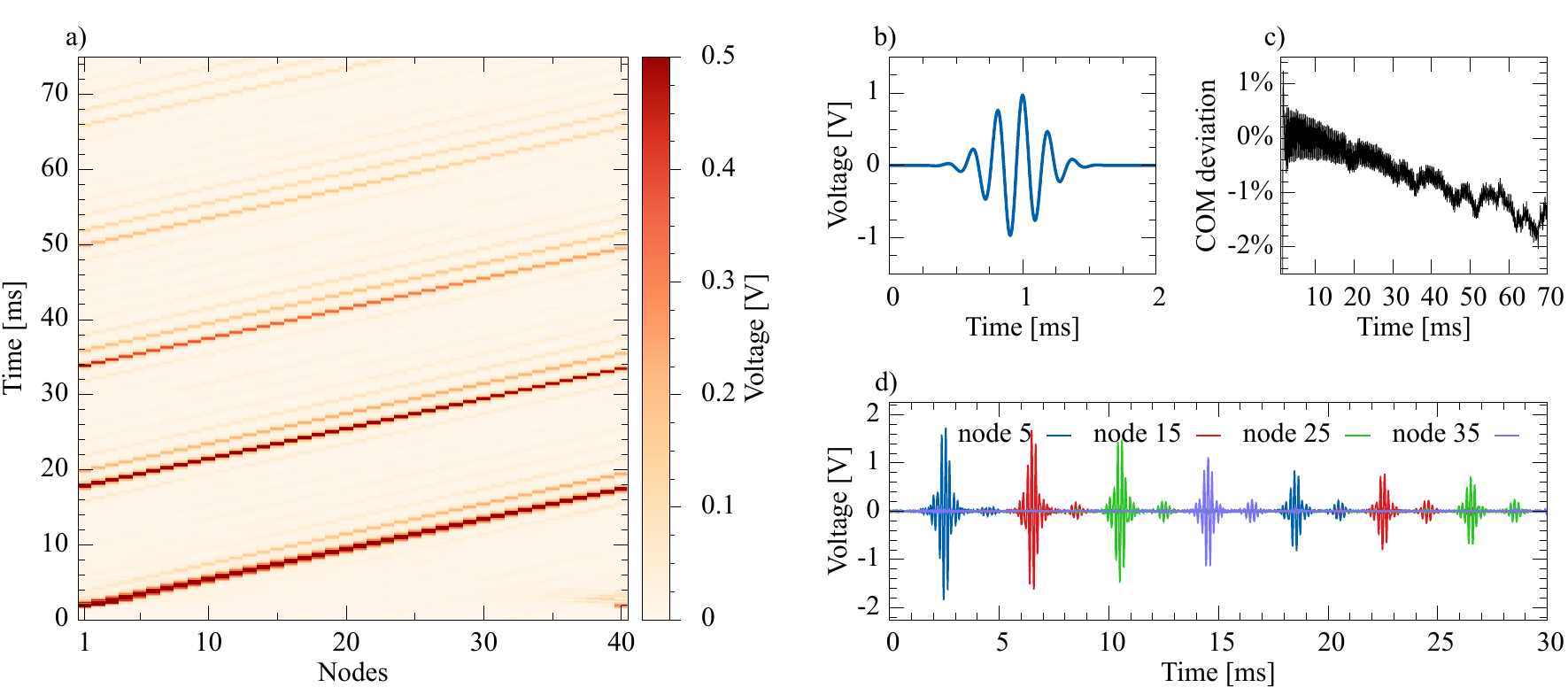}
 \caption{Adiabatic Transport in the Bulk for Periodic Boundaries. 
  a) Evolution of the voltage pulse over time. The plot shows the signal envelope extracted from measurement data using the Hilbert transform. The pumping potential shifts with a constant velocity of $2.5\,\frac{\mathrm{nodes}}{\mathrm{ms}}$. The pulse is well-localized in a minimum of the potential and propagates with a velocity proportional to the lowest band's Chern number. 
  b) Pulse used to excite the circuit for the bulk transport measurement. Its mean frequency is $5.25\,\mathrm{kHz}$, corresponding to the lowest set of bands in \hyperref[Fig:Fig3]{Fig.~2}.
 c) Relative deviation of the center of mass of the signal from the theoretical trajectory of adiabatic pumping. 
d) Voltage evolution over time for selected nodes. Some signal amplitude is observed to tunnel to the following potential minimum with each pass.}
 \label{Fig:Fig2}
\end{figure*}
}

\def \FigureThree
 {
\begin{figure}[t]
 \centering
 \includegraphics[width=\linewidth]{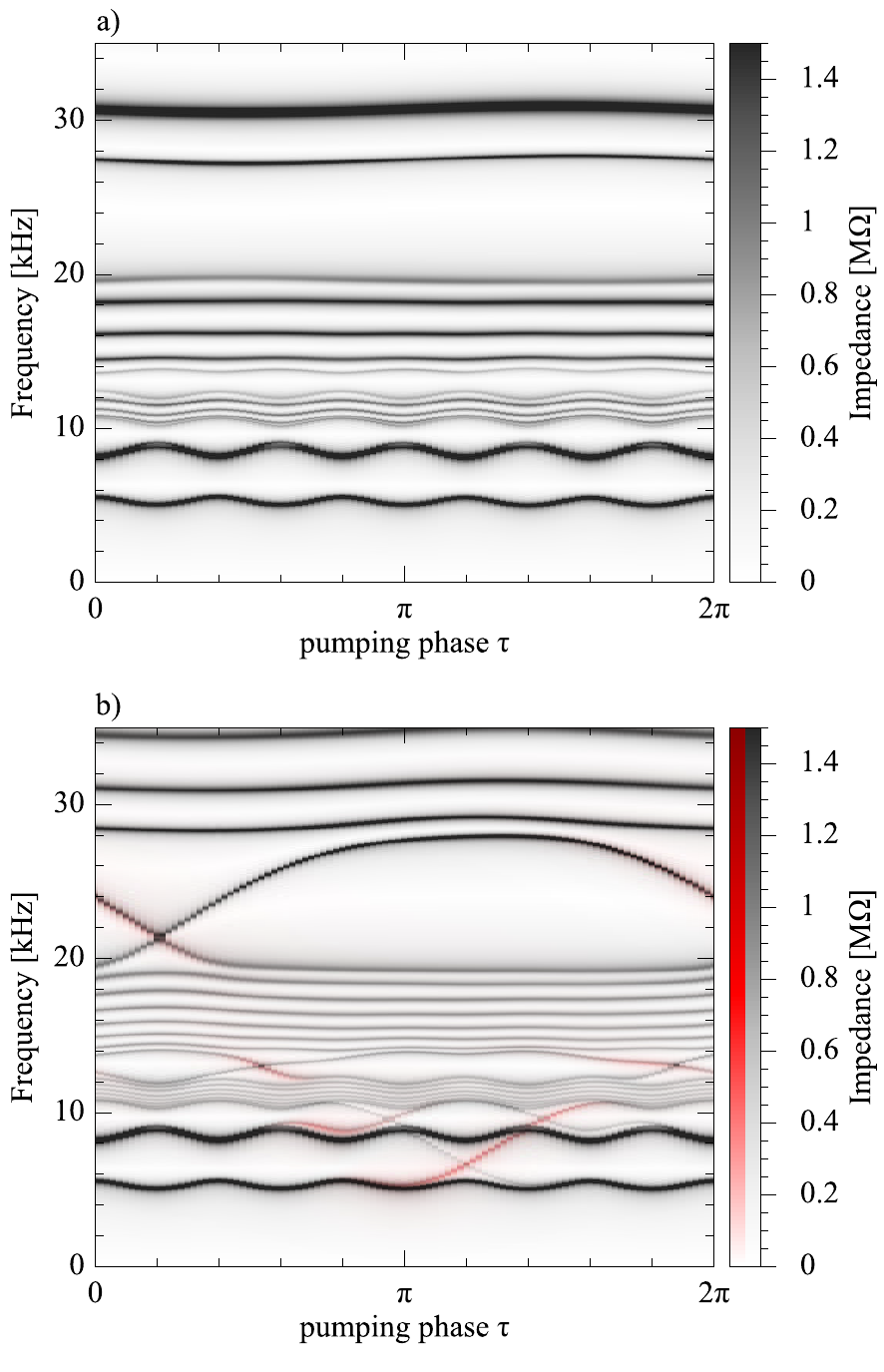}
 \caption{Measured Eigenfrequency Band Structures. Shown is the sum of
   impedances from each node to ground (absolute value) for periodic
   (a) and open boundary conditions (b), resolved over frequency and
   pumping phase. Eigenfrequencies appear as peaks in the
   impedance. The dark lines form eigenfrequency bands over pumping phase $\tau$. In (b), topological edge states emerge that cross the gaps of the band structure. 
   The impedance of the leftmost node to ground is overlaid in red, identifying edge modes localized at the left boundary.}
 \label{Fig:Fig3}
\end{figure}
}

\begin{document}


\title{Realizing efficient topological temporal pumping in electrical circuits}

 \author{Alexander Stegmaier\,\orcidlink{0000-0002-8864-5182}}
 \email[e-mail AS: ]{alexander.stegmaier@uni-wuerzburg.de}
 \affiliation{Institute for Theoretical Physics and Astrophysics, University of W\"urzburg, Am Hubland, D-97074 W\"urzburg, Germany}
 
 \author{Hauke Brand}
 \affiliation{Physikalisches Institut and R\"ontgen Research Center for Complex Material Systems, Universit\"at W\"urzburg, D-97074 W\"urzburg, Germany}
 
 \author{Stefan Imhof}
 \affiliation{Physikalisches Institut and R\"ontgen Research Center for Complex Material Systems, Universit\"at W\"urzburg, D-97074 W\"urzburg, Germany}
 
 \author{Alexander Fritzsche}
 \affiliation{Institut f\"ur Physik, Universit\"at Rostock, Albert-Einstein-Stra{\ss}e 23, 18059 Rostock}
 \affiliation{Institute for Theoretical Physics and Astrophysics, University of W\"urzburg, Am Hubland, D-97074 W\"urzburg, Germany}
 
 \author{Tobias Helbig\,\orcidlink{0000-0003-1894-0183}}
 \affiliation{Institute for Theoretical Physics and Astrophysics, University of W\"urzburg, Am Hubland, D-97074 W\"urzburg, Germany}
 
 \author{Tobias Hofmann\,\orcidlink{0000-0002-1888-9464}}
 \affiliation{Institute for Theoretical Physics and Astrophysics, University of W\"urzburg, Am Hubland, D-97074 W\"urzburg, Germany}
 
\author{Igor Boettcher\,\orcidlink{0000-0002-1634-4022}}
 \affiliation{Department of Physics, University of Alberta, Edmonton, Alberta T6G 2E1, Canada}
 \affiliation{Theoretical Physics Institute, University of Alberta, Edmonton, Alberta T6G 2E1, Canada}
 
 \author{Martin Greiter}
 \affiliation{Institute for Theoretical Physics and Astrophysics,
   University of W\"urzburg, Am Hubland, D-97074 W\"urzburg, Germany}
   
\author{Ching Hua Lee\,\orcidlink{0000-0003-0690-3238}}
\affiliation{Department of Physics, National University of Singapore,
  Singapore, 117542}
  
\author{Gaurav Bahl\,\orcidlink{0000-0001-7801-2739}}
\affiliation{Department of Mechanical Science \& Engineering,
  University of Illinois at Urbana-Champaign, Urbana, IL 61801 USA}
  
 \author{Alexander Szameit\,\orcidlink{0000-0003-0071-6941}}
 \affiliation{Institut f\"ur Physik, Universit\"at Rostock, Albert-Einstein-Stra{\ss}e 23, 18059 Rostock}
 
 \author{Tobias Kie{\ss}ling}
 \affiliation{Physikalisches Institut and R\"ontgen Research Center for Complex Material Systems, Universit\"at W\"urzburg, D-97074 W\"urzburg, Germany}

 \author{Ronny Thomale}
 \affiliation{Institute for Theoretical Physics and Astrophysics, University of W\"urzburg, Am Hubland, D-97074 W\"urzburg, Germany}
 
 \author{Lavi K. Upreti\,\orcidlink{0000-0002-1722-484X}}
 \email[e-mail LKU: ]{lavi.upreti@uni-konstanz.de}
 \affiliation{Institute for Theoretical Physics and Astrophysics, University of W\"urzburg, Am Hubland, D-97074 W\"urzburg, Germany}
 \affiliation{Fachbereich Physik, Universit\"at Konstanz, D-78457 Konstanz, Germany}

\date{\today}

\begin{abstract}
Quantized adiabatic transport can occur when a system is slowly modulated over time. In most realizations however, the efficiency of such transport is reduced by unwanted dissipation, back-scattering, and non-adiabatic effects. In this work, we realize a topological adiabatic pump in an electrical circuit network that supports remarkably stable and long-lasting pumping of a voltage signal. We further characterize the topology of our system by deducing the Chern number from the measured edge band structure. To achieve this, the experimental setup makes use of active circuit elements that act as time-variable voltage-controlled inductors.
\end{abstract}

\pacs{Valid PACS appear here}
\maketitle

\paragraph*{Introduction.---}
The Thouless pump \cite{Thouless1983} is an adiabatic charge pump, whose transport properties are characterized by its underlying topology.
The pumping process is achieved through the slow, periodic modulation of a potential, thereby inducing the transport of particles confined to a lattice despite the filled band and in absence of a net external field.
The rate of transport is quantified by a Chern number associated with the system's energy bands, which is a topological invariant of the same type as in the integer quantum Hall effect and Chern insulators \cite{Hasan2010}. However, for the Thouless pump, the Chern number is defined over a 1+1D periodic Brillouin zone constituted by one spatial dimension and time, in contrast to two spatial dimensions for the aforementioned effects.
Topological protection ensures that the quantization of charge pumping is robust and unaffected by weak disorder \cite{Niu1984}. \\
Recently, interest in the topological Thouless pump resurged following the first experimental realizations 
in ultracold atomic systems \cite{Lohse2016, Nakajima2016, Lu2016}.
Since then, implementations in a range of setups have emerged, including  photonic waveguides \cite{Ke2016, Cerjan2020, Jurgensen2021, Sun2022}, acoustic metamaterials \cite{Cheng2020, You2022}, and mechanical \cite{Grinberg2020, Xia2021} systems.
Similar platforms 
are a seed for subsequent refinement and generalizations of the principal Thouless pump motif, as they allow for devising experiments that explore variations such as disordered pumping \cite{Cerjan2020, Nakajima2021, Grinberg2020}, pumping in continuous systems \cite{Minguzzi2022}, non-linear pumping \cite{Tangpanitanon2016, Mostaan2022, Jurgensen2022}, non-Abelian pumps \cite{You2022,Brosco2021,Sun2022}, and non-adiabatic \cite{Fedorova2020} or Floquet pumping \cite{Nathan2020,Kolodrubetz2018,Upreti2020,Adiyatullin2023}. 
Realizations known to date, however, typically exhibit considerable
deviations from ideal topological pumping due to experimental
constraints such as dissipation or non-adiabaticity, so that attainable pumping distances are limited to tens of lattice sites \cite{Lohse2016,Jurgensen2021,You2022,Minguzzi2022}. 

Electrical circuit lattices are a suitable platform to emulate
topological phenomena of synthetic matter, and promise to achieve high fidelity topological pumping over long
distances. Circuit lattices have recently gained relevance as a
versatile platform for various topological systems~\cite{Simon2015, Albert2015, Lee2017, Helbig2019,Hofmann2020,Stegmaier2021,Hohmann2022,zhang2022anomalous}. The availability of quality components
provides a versatile toolbox for the implementation of a wide range of models and effects. Another asset of electrical circuits are active circuit components such as operational amplifiers (\mbox{Op-Amps}) or analog multipliers.
\mbox{Op-Amps} in particular enabled the realization of chiral edge
propagation \cite{Hofmann2019}, the non-Hermitian skin effect
\cite{Helbig2020, Zou2021}, active topological materials
\cite{Kotwal2021} and other topological phenomena \cite{Ezawa2019, Rafi-Ul-Islam2021, Olekhno2020, Xiang2020, Yao2022, Zhang2022}. 
Active elements also provide a gateway towards the implementation of temporally modulated systems in electrical circuits \cite{Darabi2020,Nagulu2022}.\\
\FigureOne

In this Letter, we present the experimental realization of a topological Thouless pump in an electrical circuit. We employ novel time modulated circuit elements to implement the Aubry-André-Harper (AAH) model \cite{Harper1955,Aubry1980}, which describes a particle on a one-dimensional chain, exposed to a periodic, time-dependent potential. 
For this, we devise circuit elements that function as a voltage controlled variable inductor, based on analog multipliers. 
The setup displays temporal pumping of a localized voltage signal that not only agrees extraordinarily well with theoretical predictions but also remains stable for a long time and over many ($> 10^2$) lattice sites. 
Through impedance measurements, we resolve the band structure of the circuit and determine the topological Chern numbers of the bands from the edge state spectrum and the transport velocity of an adiabatically pumped signal.

\paragraph*{Implementing the Thouless pump.---}
The driven Aubry-André-Harper model (AAH) is a 1D tight-binding model with a space- and time-modulated onsite potential, described by the Schr\"odinger equation
\begin{align}\label{eq:tbH_Thouless}
	\mathrm{i} \frac{\mathrm{d}}{\mathrm{d}t} \psi_{n} = j(\psi_{n+1}+\psi_{n-1}) +\lambda\cos{(n\varphi+\tau)}\,\psi_{n},
\end{align}
where $\psi_{n}$ is the wavefunction at site $n$, $j$ is the hopping amplitude, $\lambda$ is the amplitude of the onsite potential, $\varphi$ the phase difference between neighboring nodes due to spatial modulation, and $ \tau $ the pumping parameter to be modulated over time. 
The unit cell of the lattice is determined by the potential's wavelength of $2\pi/\varphi$. 
Note that if $\varphi$ is $2\pi$ times an irrational number, then the on-site term is quasiperiodic, and leads to a quasicrystal \cite{Kraus2012,Kraus2013}. For our case, we choose a rational value of $\varphi=\frac{2\pi}{5}$, corresponding to a unit cell containing five nodes. The potential is illustrated in \hyperref[Fig:Fig1]{Fig.~1a}. Our circuit setup is a chain of eight unit cells ($40$ nodes).

The hopping terms between nodes are represented in the circuit by capacitors $ C_0 $ and the on-site terms by variable inductors $ L_n(\tau)$, see \hyperref[Fig:Fig1]{Fig.\,1b}. We employ a realization of variable inductors through analog multipliers, see \hyperref[Fig:Fig1]{Fig.~1c}.
The effective inductance can be controlled through an external control voltage $V_{\textrm{c}}$. 
The resulting differential equation describing the circuit's nodal voltages $V_{m}$ for input currents $I_{n}$ at nodes $\{n\}$ is
\begin{align}\label{eq:Floquet_diff}
	\frac{\mathrm{d}}{\mathrm{d}t} I_{n}(t) =&\frac{\mathrm{d}^2}{\mathrm{d}t^2} C_0 (2\delta_{n,m}-\delta_{n+1,m}-\delta_{n-1,m})\,V_{m}(t) \nonumber \\ &+ (L_0)^{-1}[1+\lambda\,\cos{(n\varphi+\tau)}]\delta_{n,m}\, V_{m}(t) \nonumber \\
    =& \frac{\mathrm{d}^2}{\mathrm{d}t^2} \Gamma_{n m} \,V_m(t) + \Lambda_{n m}(\tau)\, V_m(t).
\end{align}

\paragraph*{Band structures.---}

\FigureThree

We first investigate the properties of the circuit for fixed values of $\tau$ in order to reveal the band topology of the system over the parameter space of $\tau$ and lattice momentum $k$.
The resonance band structure, i.e. the eigenfrequencies of the circuit, can be measured through the Green function $G(\omega)$ in the frequency domain $\omega$. For this, we Fourier transform and invert eq. (\ref{eq:Floquet_diff}), obtaining
\begin{align}
V_n(\omega) &= G_{n m}(\omega)\, I_m(\omega),\\
\text{with} \quad G(\omega) &= \left(\mathrm{i}\omega \Gamma + \frac{1}{\mathrm{i}\omega} \Lambda(\tau)\right)^{-1}.
\end{align}
Physically, the matrix elements of the Green function $G_{n m}(\omega)$
are the impedances between nodes $n$ and $m$, where a diagonal
element $G_{n n}(\omega)$ denotes the impedance from node $n$ to ground.
To measure them, we record impedance sweeps with frequencies $f =\omega/2\pi$ in the range of $0$ to $\SI{35}{\kilo\hertz}$, for different values of $\tau$ between $0$ and $2\pi$. 
Since resonances of a circuit are poles (undamped circuit) or peaks (damped circuit) in an eigenvalue $g_i(\omega)$ of $G(\omega)$, we can detect them in the trace of the Green function $\tr(G(\omega))=\sum_i g_i(\omega)$.
Its value is obtained by summing the measured impedances to ground of all nodes. 

Figure \hyperref[Fig:Fig3]{2a} shows the measured trace of the Green function for periodic boundary conditions. 
Since the circuit consists of eight unit cells of five nodes each, we expect to observe five bands, each containing eight states of different momenta. The five bands can be seen in \hyperref[Fig:Fig3]{Fig.~2a}, with the first band within the range of $5 - \SI{6}{\kilo\hertz}$, second at $8 - \SI{9}{\kilo\hertz}$, third at $10 - \SI{13}{\kilo\hertz}$, fourth at $13 - \SI{20}{\kilo\hertz}$ and the fifth band above $\SI{27}{\kilo\hertz}$. The visible sub-bands correspond to the states of different lattice momentum $k$. 
In an LC circuit, AC currents at low frequency flow predominantly through the inductive components, at high frequencies mostly through the capacitive ones. This is reflected in the observed band structure: 
The small spread within the low-frequency bands indicate a dominance of the on-site potential, realized by inductors, over the hopping terms realized by capacitors. At the same time, an oscillation of the band as a function of $\tau$ can be seen.
For high-frequency bands, this observation is reversed, with large spread between modes of differing $k$ but diminished oscillation in $\tau$, indicating a dominance of the capacitive hopping term.

We also investigate the case of open boundary conditions. Here, dangling capacitors at the end of the chain are connected to ground. The result of the open boundaries impedance measurements is shown in \hyperref[Fig:Fig3]{Fig.~2b}.
We observe the presence of edge states crossing the band gaps. The Chern numbers of the band can be determined by counting the edge states attached to each band \cite{Hatsugai1993}. 
The winding number $\nu_n$ of the gap between bands $n$ and $n+1$ is the number of ascending minus descending edge modes localized at the left boundary. We obtain winding numbers $\nu_1=1$, $\nu_2=2$, $\nu_3=-2$, and $\nu_4=-1$ by counting the left edge modes marked red in \hyperref[Fig:Fig3]{Fig.~2b}. The Chern number of each band is then the difference of winding numbers in the gap above and below $C_n = \nu_n - \nu_{n-1}$. 
For the five bands we obtain $\{1,1,-4,1,1\}$, which agrees with numerical calculations. The Chern number can also be measured from the shift in the center of mass of a pulse over one pumping period, with the same result for the lowest band as described below.

\FigureTwo

\paragraph*{Topological pumping.---}
We investigate the topological pumping of a signal pulse in a setup with periodic boundaries. 
To induce pumping, the parameter $\tau$ is modulated over time as $\tau(t)=\Omega t$ by applying oscillating control voltages $V_{\textrm{c}}(t)$.
Adiabatic evolution occurs if the pumping frequency $\Omega$ is small
compared to the resonance frequencies of the circuit. As the adiabatic
theorem is most commonly discussed in the quantum mechanical context
of the Schr\"odinger equation, we provide an analytical derivation for LC circuits in the supplemental material. We find, that the adiabatic theorem still holds but is modified by a re-scaling of voltages by the factor $\sqrt{\omega_n(0)/\omega_n(\tau)}$ depending on the resonance frequencies of the respective eigenmodes, which coincides with results previously derived for elastic materials \cite{Nassar2018}. 
The effect of temporal modulation can be understood as parallel transport of the instantaneous eigenstates along $\tau$ in parameter space.
In order to stabilize the signal over long times, a subcircuit compensating parasitic serial resistance is added to the time-dependent inductor, detailed in the supplement. 
The topological adiabatic pump transports the wavefunction of a filled band with Chern number $C_n$ by $C_n$ unit cells per pumping cycle. 

We evenly excite the states of a resonance band at a fixed frequency through the approximately uniform excitation of the lowest band by a local resonant excitation.
By its nature, a local excitation is equally distributed over all lattice momenta $k$, and since the band is approximately flat in $k$, the excitation results in approximately the same voltage amplitude in all states. This emulates a filled band in so far, that the average velocity due to dispersion always vanishes, allowing us to observe adiabatic transport with a quantized velocity \cite{Li2021}.  

We inject an AC pulse (see \hyperref[Fig:Fig2]{Fig.~3b}) with mean
frequency $f = \SI{5.25}{\kilo\hertz}$ at node one, exciting the
lowest band of the circuit. The pulse is timed with the minimum of the potential at the injection node to maximize the overlap with the lowest band's eigenstates. 
The circuit's variable inductors are driven at pumping frequency {$\Omega=\SI{500}{\hertz}$}, moving the potential across the five sites of a unit cell within one period. Accordingly, the adiabatic pumping velocity in the $n$-th band is $C_n \cdot \SI{500}{\hertz}  \cdot 5\,\mathrm{nodes} = C_n \cdot 2.5\,\frac{\mathrm{nodes}}{\mathrm{ms}}$, so it takes $\SI{16}{\milli\second}/C_n$ to traverse all $40$ nodes of the circuit. 
 
We then measure the evolution of the voltage signal in the circuit. Figure  \hyperref[Fig:Fig2]{3a} shows a density plot of the signal envelope over space (circuit nodes) and time. The signal envelope was extracted from the measurement data using the Hilbert transform as the absolute value of the analytic signal.
We observe that the signal is transported across the circuit chain, remaining localized in the potential minimum as it moves along the chain. Figure \hyperref[Fig:Fig2]{3d} displays the signal over time at different nodes. We observe that the voltage of the pulse only decays to roughly half of its amplitude after traversing the entire chain once (i.e., 40 nodes or 8 unit cells). We also observe some broadening of the pulse, tunneling into neighboring minima of the potential, preferentially the trailing ones. This effect can be best explained as a deviation from the adiabatic approximation, since dispersive effects would cause symmetric spreading with no preferred direction.
\hyperref[Fig:Fig2]{Figure 3c} shows the relative deviation of the voltage pulse's center of mass from the trajectory predicted by the lowest band's Chern number ${C_1=1}$ and pumping frequency $\Omega = \SI{500}{\hertz}$. The graph shows an excellent agreement between theory and experiment, with a relative deviation after $70$ ms of less than $2\%$. This time scale corresponds to the pulse being transported across the entire chain four times. This minimal relative deviation and the small attenuation in space and time suggest the potential of electric circuits to implement very efficient topological adiabatic pumps. 

\paragraph*{Conclusion.---}We show the experimental implementation of a topological adiabatic temporal pump induced by a parameter in the 1D electric circuit.  Previous implementations on different platforms face considerable limitations, where deviations in the center of mass of the pulse become significant even after a few pumping periods for small unit cells and system sizes. We illustrate how a quantized transport of voltage pulses in our circuit persists for time scales that are several times the pumping period, which is much longer than in previous realizations. We also quantify the small attenuation in the center of mass of the pulse. 
Moreover, the distance between circuit nodes in the system is immaterial and
not related to any physical length scale; hence the entire
implementation can be miniaturized or expanded significantly. The
propagation velocity of the pulse with respect to other parametric
variations can likewise be efficiently controlled. Such flexibility
and easy implementation with inexpensive components highlight the
potential for practical applications, where the physical length and
time scales would then be matched to the problem at hand.

\paragraph*{Acknowledgement.---}
The work is funded by the Deutsche Forschungsgemeinschaft (DFG, German Research Foundation) through Project-ID 258499086 - SFB 1170 and through the W\"urzburg-Dresden Cluster of Excellence on Complexity and Topology in Quantum Matter -- \textit{ct.qmat} Project-ID 390858490 - EXC 2147. T.He. was supported by a Ph.D. scholarship of the German Academic Scholarship Foundation. IB acknowledges support from the University of Alberta startup fund UOFAB Startup Boettcher and the Natural Sciences and Engineering Research Council of Canada (NSERC) Discovery Grants RGPIN-2021-02534 and DGECR2021-00043. LKU acknowledges support from the SNF grant 13947622-FP476/22Zilberberg.

\end{document}



\title{Supplemental material to:\\ Realizing efficient topological temporal pumping in electrical circuits}

\author{Alexander Stegmaier}
 \affiliation{Institute for Theoretical Physics and Astrophysics, University of W\"urzburg, Am Hubland, D-97074 W\"urzburg, Germany}
  \author{Hauke Brand}
 \affiliation{Physikalisches Institut and R\"ontgen Research Center for Complex Material Systems, Universit\"at W\"urzburg, D-97074 W\"urzburg, Germany}
 \author{Stefan Imhof}
 \affiliation{Physikalisches Institut and R\"ontgen Research Center for Complex Material Systems, Universit\"at W\"urzburg, D-97074 W\"urzburg, Germany}
 \author{Alexander Fritzsche}
 \affiliation{Institut f\"ur Physik, Universit\"at Rostock, Albert-Einstein-Stra{\ss}e 23, 18059 Rostock}
 \affiliation{Institute for Theoretical Physics and Astrophysics, University of W\"urzburg, Am Hubland, D-97074 W\"urzburg, Germany}
\author{Tobias Helbig}
 \affiliation{Institute for Theoretical Physics and Astrophysics, University of W\"urzburg, Am Hubland, D-97074 W\"urzburg, Germany}
 \author{Tobias Hofmann}
 \affiliation{Institute for Theoretical Physics and Astrophysics, University of W\"urzburg, Am Hubland, D-97074 W\"urzburg, Germany}
 \author{Igor Boettcher}
 \affiliation{Department of Physics, University of Alberta, Edmonton, Alberta T6G 2E1, Canada}
\affiliation{Theoretical Physics Institute, University of Alberta, Edmonton, Alberta T6G 2E1, Canada}
 \author{Martin Greiter}
 \affiliation{Institute for Theoretical Physics and Astrophysics,
   University of W\"urzburg, Am Hubland, D-97074 W\"urzburg, Germany}
\author{Ching Hua Lee}
\affiliation{Department of Physics, National University of Singapore,
  Singapore, 117542}
\author{Gaurav Bahl}
\affiliation{Department of Mechanical Science \& Engineering,
  University of Illinois at Urbana-Champaign, Urbana, IL 61801 USA}
  \author{Alexander Szameit}
 \affiliation{Institut f\"ur Physik, Universit\"at Rostock, Albert-Einstein-Stra{\ss}e 23, 18059 Rostock}
 \author{Tobias Kie{\ss}ling}
\affiliation{Physikalisches Institut and R\"ontgen Research Center for Complex Material Systems, Universit\"at W\"urzburg, D-97074 W\"urzburg, Germany}
 \author{Ronny Thomale}
 \affiliation{Institute for Theoretical Physics and Astrophysics, University of W\"urzburg, Am Hubland, D-97074 W\"urzburg, Germany}
 \author{Lavi K. Upreti}
 \affiliation{Institute for Theoretical Physics and Astrophysics, University of W\"urzburg, Am Hubland, D-97074 W\"urzburg, Germany}
 \affiliation{Fachbereich Physik, Universit\"at Konstanz, D-78457 Konstanz, Germany}
\date{\today}

\pacs{Valid PACS appear here}
\maketitle

\onecolumngrid
\appendix
\renewcommand{\thefigure}{S\arabic{figure}}
\setcounter{figure}{0}

\section{A. Theory}

\subsection{Topology of the Aubry-Andr\'e-Harper model}

The Aubry-Andr\'e-Harper (AAH) model is a 1D hopping chain with a cosine-shaped, modulated on-site potential. Its Hamiltonian takes the form
\begin{align}
    H(\tau) = \frac{1}{2}\sum_i \left(-t\, c_i^\dagger c_{i+1} + \lambda \cos{(\tau - i \varphi)} c_i^\dagger c_i + \text{h.c.}\right)
\end{align}
The potential is shifted by a phase $\varphi$ between neighboring nodes. If $\varphi$ is a fraction of $2\pi$, $\varphi = 2\pi/N$, then the model forms a lattice with a periodicity of $N$ sites. The corresponding Bloch Hamiltonian is 
\begin{equation}
    h(k,\tau) = 
    \begin{pmatrix}
    \lambda \cos{(\tau)} & -t & \cdots& -t \,\mathrm{e}^{-\mathrm{i} k}\\
    -t & \lambda \cos{(\tau-2\pi/N)} & -t & \vdots \\
    \vdots & -t & \ddots & -t\\
     -t \,\mathrm{e}^{\mathrm{i}k} &\cdots & -t& \lambda \cos{(\tau - 2\pi (N-1)/N)}
    \end{pmatrix}.
    \end{equation}
We consider the dynamics of an initially localized state in the $n^{\text{th}}$ band under adiabatic variation of $\tau$. In this scenario, the AAH model acts as a Thouless pump, transporting the state by $C_n$ unit cells, with $C_n$ the Chern number of the $n^{\text{th}}$ band over to the 2D parameter space spanned by lattice momentum $k$ and pumping phase $\tau$. The AAH model is invariant under a combined translation of one site and a shift in the pumping phase $\tau$ by $\varphi$. Semi-classically, it can be argued that states bound to the minima (or maxima) of the pumping potential should be transported by one unit cell per pumping period, yielding a Chern number of one for the corresponding bands. This argument can be formalized by investigating the Berry curvature over the $\{k, \tau\}$ parameter space. The Chern number of the $n^{\text{th}}$ band is given by the integral of the Berry curvature
\begin{align}
    C_n = \frac{1}{2\pi\mathrm{i}} \iint \mathrm{d}k\, \mathrm{d}\tau\, \left[\partial_\tau (\bm{\psi}_n^\dagger \partial_k\bm{\psi}_n) - \partial_k (\bm{\psi}_n^\dagger \partial_\tau \bm{\psi}_n)\right],
\end{align}
with $\bm{\psi}_n(k,\tau)$ being the $n^{\text{th}}$ eigenvector of $h(k, \tau)$. Let us investigate the first term of the integral, $\partial_\tau \bm{\psi}_n^\dagger \partial_k\bm{\psi}_n$. Assuming a smooth gauge can be chosen within the stripes $0 \leq k < 2\pi,\; (i-1)\varphi\leq \tau < i\varphi$, it can be re-written as
\begin{align}
    \iint \mathrm{d}k\, \mathrm{d}\tau\, \partial_\tau (\bm{\psi}_n^\dagger \partial_k\bm{\psi}_n) 
    &= \int \mathrm{d}k \left(\left[\bm{\psi}_n^\dagger \partial_k\bm{\psi}_n\right]_{\tau=0}^{\varphi} + \left[\bm{\psi}_n^\dagger \partial_k\bm{\psi}_n\right]_{\tau=\varphi}^{2\varphi} + \ldots + \left[\bm{\psi}_n^\dagger \partial_k\bm{\psi}_n\right]_{\tau=2\pi-\varphi}^{2\pi}\right)\\
    &=\sum_{i=1}^N \int \mathrm{d}k \left[ \bm{\psi}_n^\dagger \partial_k \bm{\psi}_n \right]_{\tau=(i-1)\varphi}^{i\varphi}.
\end{align}
Let the eigenvector $\bm{\psi}_n(k,\tau)$ at $\tau=0$ be $\bm{\psi}_n(k,0) = (a_1,a_2, \ldots, a_N)^\intercal$. At $\tau=\varphi$, the Hamiltonian is shifted by one site compared to $\tau=0$. Accordingly, the eigenvectors become $\bm{\psi}_n(k,\varphi) = (\mathrm{e}^{\mathrm{i}k} a_N ,a_1,a_2, \ldots, a_{N-1})^\intercal$. Generally, the eigenvectors for $\tau = i \varphi$ are $\bm{\psi}_n(k,i\varphi) = (\mathrm{e}^{\mathrm{i}k} a_{N-i+1}, \ldots, \mathrm{e}^{\mathrm{i}k} a_N, a_1, \ldots, a_{N-i})^\intercal$. Substituting back to the integral, we obtain
\begin{align}
    \sum_{i=1}^N \int \mathrm{d}k \left[ \bm{\psi}_n^\dagger \partial_k \bm{\psi}_n \right]_{\tau=(i-1)\varphi}^{i\varphi}
    =& \sum_{i=1}^N \int \mathrm{d}k\; \left[\bm{\psi}_n^\dagger(k,i\varphi) \partial_k \bm{\psi}_n(k,i\varphi)-\bm{\psi}_n^\dagger(k,(i-1)\varphi) \partial_k \bm{\psi}_n(k,(i-1)\varphi)\right] \\
    =& \sum_{i}^N \int \mathrm{d}k\; \mathrm{i}|a_i|^2\\
    =& \int_0^{2\pi}\mathrm{d}k\; \mathrm{i} = 2\pi\mathrm{i}.
\end{align}

This term contributes $+1$ to the Chern number of the respective band. While this calculation does not generate a definitive Chern number for any band, since the second term of the Berry curvature was neglected and assumptions about the smoothness of the gauge need to be made,  it shows how the shifting potential biases the bands of the AAH model towards a Chern number of $1$. For a finite number of bands, the total Chern number of a system is necessarily zero, so some bands must always violate this result to compensate, i.e. there must always be some left-moving modes compensating the right-moving ones.

Another feature that makes the AAH model well suited for the demonstration of topological pumping is the strong localization of low energy modes. For sufficiently small $\varphi$, these essentially correspond to bound states in the wells of the cosine potential that are only weakly coupled to neighboring wells. 
As a result, low energy bands are relatively flat, so that a pulse can stay localized over many pumping cycles.

\subsection{Topological states as resonances of an L-C network} \label{app:dynamics}

The nodal voltages of an L-C network are described by the system of second-order differential equations
\begin{align}
    \Gamma \frac{\mathrm{d}^2}{\mathrm{dt}^2} \bm{V}(t) + \Lambda \bm{V}(t) = \frac{\mathrm{d}}{\mathrm{d}t} \bm{I}(t),
\end{align}
where $\bm{V}$ is the vector of nodal voltages and $\bm{I}$ the vector of all external currents flowing into the nodes. 
$\Gamma$ and $\Lambda$ are the admittance matrices corresponding to the capacitive and inductive circuit elements respectively. 
In the frequency domain, the according equation is
\begin{align}
(-\omega^2 \Gamma + \Lambda) \bm{V}(\omega)=\mathrm{i}\omega J(\omega) \bm{V}(\omega) = \bm{I}(\omega),    
\end{align}
where $J(\omega)$ we call the circuit Laplacian. We use this picture of the circuit Laplacian to relate the properties of the quantum mechanical system to the electrical one, by analogy between $J$ and Hamiltonian $H$.
To calculate the eigenmodes and -frequencies of a circuit, we consider the homogeneous case $\bm{I}=\bm{0}$ and use an exponential ansatz for the eigenmodes $\bm{V}_n(t) = \mathrm{e}^{\mathrm{i}\omega_n t} \bm{V}_n$ to obtain the equation
\begin{align}
    \left(\Gamma^{-1}\Lambda\right) \bm{V}_n = \omega_n^2 \bm{V}_n.
\end{align}
While this approach provides an efficient way to calculate the eigenmodes of the circuit, it does not explain how dynamical eigenstates of the circuit and of the quantum-mechanical Hamiltonian relate. For this, another equivalent defining relation for eigenmodes can be used,
\begin{align}
    J(\omega_n)\bm{V}_n=\bm{0}.
\end{align}
This relation shows that dynamical modes emerge from the spectrum and eigenvectors of the circuit Laplacian $J(\omega)$ are the eigenpairs at the roots of the admittance eigenvalues $j_n(\omega)$. Consider a set of eigenvectors \mbox{$\{\,\bm{V}_n(k,\omega(k))\;\vert\; k\in [0,2\pi[ \, \}$} of the Laplacian of a lattice model. The states of the admittance band at some frequency $\omega$ can be recovered by setting $\omega(k) = \omega$, while the states of the eigenfrequency band is obtained by setting $\omega(k)$ to the dispersion relation $\omega_n(k)$.
This implies that, if no band-crossing occurs in the admittance band structure within the band-width of $\omega_n(k)$, the admittance eigenmodes of the Laplacian fixed frequency can be related to the eigenfrequency band structure by a continuous deformation.
Accordingly, the band topology of the dynamical eigenstates is equivalent to that of the corresponding admittance eigenstates at a suitably chosen frequency $\omega_0$, which themselves are analogous to that of the respective Hamiltonian $H$ after which $J(\omega_0)$ was modeled.

\subsection{Adiabatic evolution in a modulated L-C electric circuit} \label{app:adiabatic}
To re-derive the adiabatic theorem for an L-C circuit, we first re-express the circuit's differential equation in the canonical form
\begin{align}
  \mathrm{i} \frac{\mathrm{d}}{\mathrm{d}t} 
  \begin{pmatrix}
    \dot{\bm{V}}\\
    \bm{V}
    \end{pmatrix} =
    \mathrm{i}\begin{pmatrix}
    0 & -\Gamma^{-1}\Lambda\\
    \mathbb{1} & 0
    \end{pmatrix} 
    \begin{pmatrix}
    \dot{\bm{V}}\\
    \bm{V}
    \end{pmatrix}.
\end{align}
This differential equation is equivalent to a non-Hermitian Schr\"odinger equation with 
\begin{align}
    H= \mathrm{i}
    \begin{pmatrix}
    0 & -\Gamma^{-1}\Lambda\\
    \mathbb{1} & 0
    \end{pmatrix} .
\end{align} 
The matrix $\Gamma^{-1}\Lambda$ has eigenvalues and -vectors $\omega_n^2$ and $\bm{V}_n$, so $\Gamma^{-1}\Lambda \bm{V}_n = \omega_n^2 \bm{V}_n.$ 
In the context of an L-C circuit with positive, reciprocal capacitive and inductive couplings (such as any circuit composed of conventional passive elements), the matrices $\Gamma$ and $\Lambda$ are real-valued, symmetric, positive semi-definite and diagonally dominant, so their eigenvalues are real and positive. 
Since $\Gamma$ is symmetric with positive eigenvalues, so is $\Gamma^{-1}$. The product $\Gamma^{-1}\Lambda$ on the other hand is generally not symmetric, since the two factor matrices generally do not commute. However, direct calculation shows that it remains positive (semi-) definite, so all its eigenvalues are real and positive. This implies their square roots $\pm \omega_n$ are real-valued as well. The eigenvectors to $(\Gamma^{-1}\Lambda)^\intercal=\Lambda\Gamma^{-1}$ form the set of left eigenvectors $W_n$ that form a dual basis with the $V_n$, so that $W_n^\intercal V_m = \delta_{nm}$.\\
A set of eigenvectors $\bm{\Psi}_{n \sigma}$ to $H$, with $\sigma \in \{+, -\}$ and eigenvalues $\omega_{n\sigma}=\sigma \omega_n$ (where we choose $+\omega_n$ to always be the positive root of the corresponding eigenvalue of $\Gamma^{-1}\Lambda$) can be constructed from the $\bm{V}_n$ as 
\begin{align}
    \bm{\Psi}_{n \sigma} = \begin{pmatrix}
    - \sigma \mathrm{i} \omega_n \bm{V}_n\\
    \bm{V}_n
    \end{pmatrix}.
\end{align}
The left eigenvectors of $H$ are given by
\begin{align}
   \bm{\Phi}_{n \sigma} = \frac{1}{2\sigma \mathrm{i}\omega_n}
    \begin{pmatrix}
    \bm{W}_n\\ 
    \sigma \mathrm{i}\omega_n \bm{W}_n
    \end{pmatrix}.
\end{align}
A quick calculation confirms that these form the dual basis to $\bm{\Psi}_{n\sigma}$, namely
\begin{align}
    \bm{\Phi}_{m\rho}^\dagger \bm{\Psi}_{n\sigma} 
    &=  \frac{1}{-2\sigma \mathrm{i}\omega_n} \begin{pmatrix}
    \bm{W}_m^\dagger, - \rho \mathrm{i}\omega_m \bm{W}_m^\dagger
    \end{pmatrix}
    \begin{pmatrix}
    - \sigma \mathrm{i} \omega_n \bm{V}_n\\
    \bm{V}_n
    \end{pmatrix}\\
    &= \frac{1}{-2\sigma \mathrm{i}\omega_n}\left( -\sigma \mathrm{i} \omega_n \delta_{mn} - \rho \mathrm{i} \omega_m \delta_{mn}\right)\\
    &= \frac{-\sigma \mathrm{i}\omega_n}{-2\sigma \mathrm{i}\omega_n} (1 + \rho \sigma)\delta_{mn}\\
    &= \delta_{\rho \sigma} \delta_{m n}.
\end{align}
A general state of the circuit is described as a linear combination of the eigenstates $\bm{\Psi}(t) = \sum_{n\sigma} c_{n\sigma}(t) \bm{\Psi}_{n\sigma}(t)$. From the reality constraint of $\bm{V}$ and $\dot{\bm{V}}$, we obtain that the coefficients corresponding to the same $n$ must be complex conjugates of each other, $c_{n-}=c_{n+}^\ast$. The voltages of the system are then simply the second component of $\bm{\Psi}(t)$, so that $\bm{V}(t) = \sum_{n\sigma} c_{n\sigma}(t) \bm{V}_{n}(t)$. Using these conventions, we can use the same set of coefficients $c_{n\sigma}$ to describe both the system's state vector $\bm{\Psi}$ and voltage vector $\bm{V}$.
To derive the adiabatic theorem for slowly modulated $L-C$ circuits, we start from the Schr\"odinger equation projected onto a left eigenstate and simplify from there. We have
\begin{align}
    \bm{\Phi}_{m\rho}^\dagger \, H \sum_{n \sigma} c_{n\sigma} \bm{\Psi}_{n\sigma}&=\bm{\Phi}_{m\rho}^\dagger\, \mathrm{i}\frac{\mathrm{d}}{\mathrm{d}t} \sum_{n \sigma} c_{n\sigma} \bm{\Psi}_{n\sigma}\\
    \bm{\Phi}_{m\rho}^\dagger \, \sum_{n \sigma} \sigma \omega_n c_{n\sigma} \bm{\Psi}_{n\sigma}&=\mathrm{i} \bm{\Phi}_{m\rho}^\dagger\, \sum_{n \sigma} \left(\dot{c}_{n\sigma} \bm{\Psi}_{n\sigma} + c_{n\sigma} \dot{\bm{\Psi}}_{n\sigma}\right)\\
     \rho \omega_m c_{m\rho}&=\mathrm{i} \dot{c}_{m\rho} +  \mathrm{i} \sum_{n \sigma}  c_{n\sigma} \bm{\Phi}_{m\rho}^\dagger \dot{\bm{\Psi}}_{n\sigma}\\
    &=\mathrm{i} \dot{c}_{m\rho} +  
    \mathrm{i} \sum_{n \sigma}  c_{n\sigma} \frac{1}{-2\rho \mathrm{i}\omega_m} \begin{pmatrix}
    \bm{W}_m^\dagger, - \rho \mathrm{i}\omega_m \bm{W}_m^\dagger
    \end{pmatrix}
    \begin{pmatrix}
    -\sigma \mathrm{i} (\dot{\omega}_n \bm{V}_n + \omega_n \dot{\bm{V}}_n)\\
    \dot{\bm{V}}_n
    \end{pmatrix}\\
    &=\mathrm{i} \dot{c}_{m\rho} +  
    \mathrm{i} \sum_{n \sigma}  c_{n\sigma} \frac{1}{-2\rho \mathrm{i}\omega_m} \left( -\sigma \mathrm{i} \dot{\omega}_n \delta_{mn} -\sigma \mathrm{i} \omega_n \bm{W}_m^\dagger \dot{\bm{V}}_n -\rho \mathrm{i} \omega_m \bm{W}_m^\dagger \dot{\bm{V}}_n \right)\\
    &=\mathrm{i} \dot{c}_{m\rho} +  
    \mathrm{i} \sum_{n \sigma}  c_{n\sigma} \frac{1}{2} \left(\sigma\rho \frac{\dot{\omega}_n}{\omega_m} \delta_{mn} + (\sigma\rho \frac{\omega_n}{\omega_m} + 1) \bm{W}_m^\dagger \dot{\bm{V}}_n \right)\\
    &=\mathrm{i} \dot{c}_{m\rho} + \mathrm{i}\rho \frac{\dot{\omega}_m}{\omega_m} \frac{c_{m+}-c_{m-}}{2}   
    + \mathrm{i} \sum_{n \sigma}  c_{n\sigma} \frac{1}{2} \left(\rho \sigma \frac{\omega_n}{\omega_m} + 1 \right) \bm{W}_m^\dagger \dot{\bm{V}}_n\\
    &=\mathrm{i} \dot{c}_{m\rho} - \frac{\dot{\omega}_m}{\omega_m} \Im{c_{m+}}
    +\mathrm{i} c_{m \rho} \bm{W}_m^\dagger \dot{\bm{V}}_m
    + \mathrm{i} \sum_{n\neq m \sigma}  c_{n\sigma} \frac{1}{2} \left(\rho \sigma\frac{\omega_n}{\omega_m} + 1 \right) \bm{W}_m^\dagger \dot{\bm{V}}_n\\
    &=\mathrm{i} \dot{c}_{m\rho} - \frac{\dot{\omega}_m}{\omega_m} \Im{c_{m+}}
    +\mathrm{i} c_{m \rho} \bm{W}_m^\dagger \dot{\bm{V}}_m
    + \mathrm{i} \sum_{n\neq m \sigma}  c_{n\sigma} \frac{1}{2} \left(1+ \rho\sigma +\rho\sigma(\frac{\omega_n}{\omega_m}-1) \right) \bm{W}_m^\dagger \dot{\bm{V}}_n\\
    &=\mathrm{i} \dot{c}_{m\rho} - \frac{\dot{\omega}_m}{\omega_m} \Im{c_{m+}}
    +\mathrm{i} c_{m \rho} \bm{W}_m^\dagger \dot{\bm{V}}_m
    + \mathrm{i} \sum_{n\neq m}  c_{n \rho} \bm{W}_m^\dagger \dot{\bm{V}}_n  - \sum_{n\neq m} \rho \Im{c_{n +}} (\frac{\omega_n}{\omega_m} -1)\bm{W}_m^\dagger \dot{\bm{V}}_n.
\end{align}
We arrive at
\begin{align}
    \dot{c}_{m\rho} + \left(\mathrm{i} \rho \omega_m + \bm{W}_m^\dagger \dot{\bm{V}}_m \right) c_{m\rho} 
    &= -\sum_{n\neq m}  c_{n \rho} \bm{W}_m^\dagger \dot{\bm{V}}_n - \mathrm{i} \sum_{n} \Im{c_{n +}} \left( \frac{\dot{\omega}_m}{\omega_m} \delta_{nm} +\rho (\frac{\omega_n}{\omega_m} -1)\bm{W}_m^\dagger \dot{\bm{V}}_n\right).
\end{align}
This can be simplified further by setting $\rho=+1$, removing the redundant negative frequency case that is related to the positive frequency coefficients by complex conjugation. Suppressing the now redundant second index, the equation then reads
\begin{align}
    \dot{c}_{m} + \left(\mathrm{i} \omega_m + \bm{W}_m^\dagger \dot{\bm{V}}_m \right) c_{m} 
    &= -\sum_{n\neq m}  c_{n} \bm{W}_m^\dagger \dot{\bm{V}}_n - \mathrm{i} \sum_{n} \Im{c_{n}} \left( \frac{\dot{\omega}_m}{\omega_m} \delta_{nm} + (\frac{\omega_n}{\omega_m} -1)\bm{W}_m^\dagger \dot{\bm{V}}_n\right). \label{cm_EOM}
\end{align}
The terms on the left hand side correspond to the dynamical evolution of the quasistatic eigenstates and the geometric component of the transport that results in the Berry phase, identical to the known quantum mechanical case.
On the right-hand side of the equation, the first term is analogous to the quantum mechanical case (and neglected in the conventional adiabatic approximation). However, an additional term proportional to the imaginary part of the coefficients $c_{n}$ appears. 
To investigate which of these terms can be neglected under adiabatic evolution, we substitute the coefficients $c_n(t)$ by $e^{-\mathrm{i}\int^t \omega_n(t') \mathrm{d}t'} \tilde{c}_n(t)$. This cancels the dynamical term $\mathrm{i}\omega_n c_n$ on the left-hand side of eq. (\ref{cm_EOM}) and after multiplying the equation by $e^{\mathrm{i}\int^t \omega_m(t') \mathrm{d}t'}$ we obtain
\begin{align}
    \dot{\tilde{c}}_{m} + \bm{W}_m^\dagger \dot{\bm{V}}_m \tilde{c}_{m} 
    = -\sum_{n\neq m} e^{-\mathrm{i}\int^t (\omega_n-\omega_m) \mathrm{d}t'}  \tilde{c}_{n} \bm{W}_m^\dagger \dot{\bm{V}}_n - \mathrm{i} \sum_{n} \Im{e^{-\mathrm{i}\int^t \omega_n \mathrm{d}t'}\tilde{c}_{n}} e^{\mathrm{i}\int^t \omega_m \mathrm{d}t'} \left( \frac{\dot{\omega}_m}{\omega_m} \delta_{nm} + (\frac{\omega_n}{\omega_m} -1)\bm{W}_m^\dagger \dot{\bm{V}}_n\right)\\
    = -\sum_{n\neq m} e^{-\mathrm{i}\int^t (\omega_n-\omega_m) \mathrm{d}t'}  \tilde{c}_{n} \bm{W}_m^\dagger \dot{\bm{V}}_n -\sum_{n} \frac{e^{-\mathrm{i}\int^t (\omega_n-\omega_m) \mathrm{d}t'}\tilde{c}_{n}-e^{\mathrm{i}\int^t (\omega_n+\omega_m) \mathrm{d}t'}\tilde{c}_{n}^*}{2} \left( \frac{\dot{\omega}_m}{\omega_m} \delta_{nm} + (\frac{\omega_n}{\omega_m} -1)\bm{W}_m^\dagger \dot{\bm{V}}_n\right).
\end{align}
We see that all terms with $n\neq m$ on the right side oscillate with a finite frequency. If adiabatic evolution is assumed, these terms vanish since their contribution to $c(\tau)$ is of order $\Omega / (\omega_n-\omega_m)$. Only one of the right-hand side terms remains,
\begin{align}
  \dot{\tilde{c}}_{m} + \bm{W}_m^\dagger \dot{\bm{V}}_m \tilde{c}_{m}&=-\frac{1}{2} \frac{\dot{\omega}_m}{\omega_m} \tilde{c}_{m}\\
  \iff \dot{\tilde{c}}_{m} &= -\left(\bm{W}_m^\dagger \dot{\bm{V}}_m +\omega_m^{-1} \dot{\omega}_m /2 \right) \tilde{c}_m.
\end{align}
The term $\bm{W}_m^\dagger \dot{\bm{V}}_m$ is the Berry phase of a non-Hermitian system, analogous to the quantum mechanical case of a non-Hermitian Hamiltonian.                       ) 
The other term $\omega_m^{-1} \dot{\omega}_m /2$ has no direct analogue in quantum mechanics. Both these terms induce a purely geometric evolution in parameter space along the curve $C:\; [t_0,t]\rightarrow \mathbb{R}^N$, parameterized by $\bm{R}(t)$, via 
\begin{align}
    \dot{\tilde{c}}_m = (\bm{\nabla}_{\!R}\, c_m) \frac{\mathrm{d}\bm{R}}{\mathrm{d}t} &= -\left(\bm{W}_m^\dagger \bm{\nabla}_{\!R}\, \bm{V}_m +\omega_m^{-1} \bm{\nabla}_{\!R}\, \omega_m /2 \right) \frac{\mathrm{d}\bm{R}}{\mathrm{d}t}\, \tilde{c}_m\\
    \bm{\nabla}_{\!R}\, \tilde{c}_m &= -\left(\bm{W}_m^\dagger \bm{\nabla}_{\!R}\, \bm{V}_m +\omega_m^{-1} \bm{\nabla}_{\!R}\, \omega_m /2 \right) \tilde{c}_m\\
    \implies \tilde{c}_m(\bm{R}) &= \exp\left({\int_C -\left(\bm{W}_m^\dagger \bm{\nabla}_{\!R}\, \bm{V}_m +\omega_m^{-1} \bm{\nabla}_{\!R}\, \omega_m /2 \right) \mathrm{d}\bm{R}}\right) \tilde{c}_m(\bm{R}_0)\\
    &= \exp\left({\int_C -\bm{W}_m^\dagger \bm{\nabla}_{\!R}\, \bm{V}_m \mathrm{d}\bm{R}} - \ln{\left(\frac{\omega_m(\bm{R})}{\omega_m(\bm{R}_0)}\right)}/2 \right) \tilde{c}_m(\bm{R}_0)\\
    &= \sqrt{\frac{\omega_m(\bm{R}_0)}{\omega_m(\bm{R})}}\, e^{\mathrm{i} \gamma_m[C]}\,  \tilde{c}_m(\bm{R}_0).
\end{align}
Here we see that ultimately, the difference between adiabatic evolution in quantum mechanics and electrical L-C circuits is a re-scaling of the voltage amplitude by $\sqrt{\frac{\omega_m(\bm{R}_0)}{\omega_m(\bm{R})}}$, the square root of the ratio of initial and final eigenfrequency of the respective state.

\section{B. Circuit setup}

\begin{figure}
    \centering
    \includegraphics{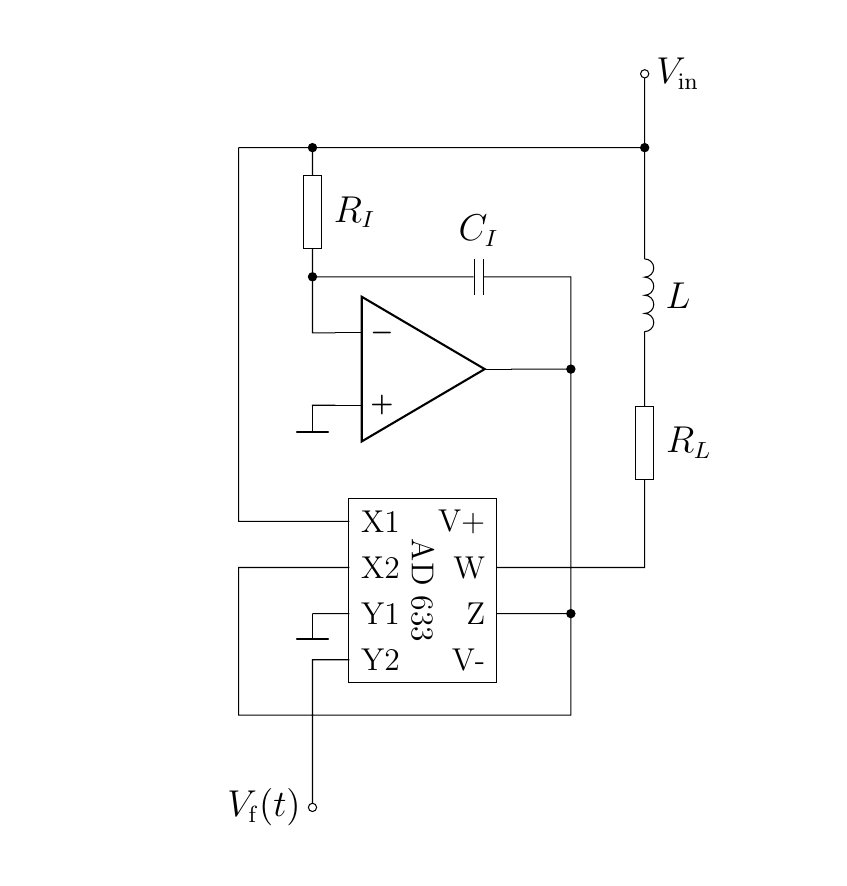}
    \caption{Floquet inductor including the serial resistance compensation subcircuit. Not shown: Voltage divider at the output of the operational amplifier for fine-tuning.}
    \label{fig:floq_inductor}
\end{figure}

\subsection{Floquet element with loss compensation}\label{app:losscomp}

The time variable inductors used in our experimental setup are implemented using AD633 analog multipliers. The multipliers have voltage inputs named X1, X2, Y1, Y2 and Z, and voltage output W. The ideal output voltage is given by 
\begin{align}
\mathrm{W}= \frac{(\mathrm{X}1-\mathrm{X}2)(\mathrm{Y}1-\mathrm{Y}2)}{10\,\mathrm{V}}+\mathrm{Z}.
\end{align}
To create the effective variable inductor as shown in Fig.\,\ref{fig:floq_inductor}, the voltage of the connected node is fed into X1, and control voltage $V_\mathrm{f}$ to Y2. Then the voltage across inductor $L$ connected between the connected node and W is $V_\mathrm{in} + V_\mathrm{in}\frac{V_\mathrm{f}}{10\,\mathrm{V}} = (1+\frac{V_\mathrm{f}}{10\,\mathrm{V}})\,V_\mathrm{in}.$ The current through the inductor is then 
\begin{align}
I = \int \mathrm{d}t \; (1+\frac{V_\mathrm{f}}{10\,\mathrm{V}})\,L^{-1}\; V_\mathrm{in}.
\end{align}

\begin{figure}
     \centering    \includegraphics{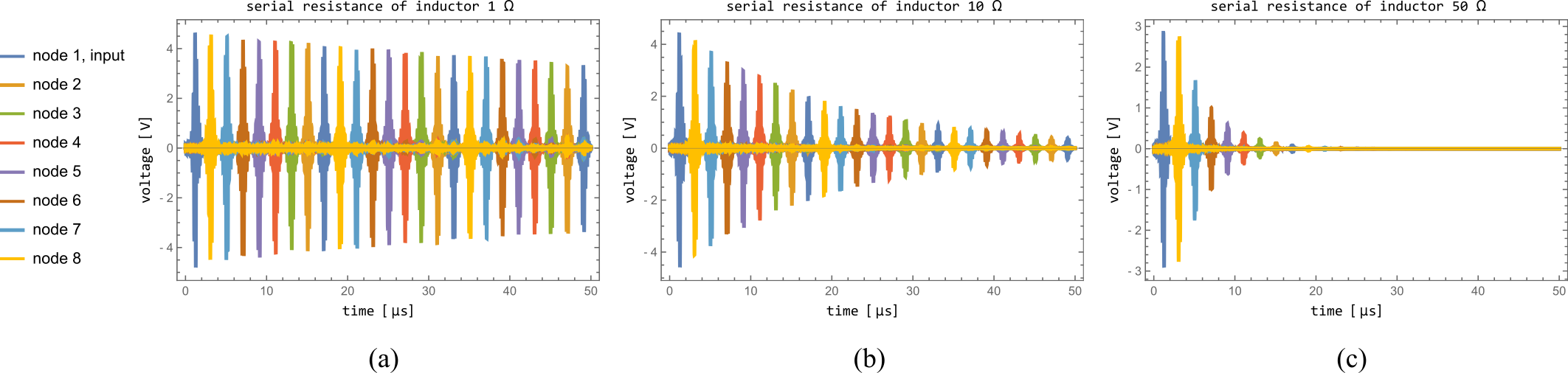}
    \caption{Influence of the compensation of the serial resistance of the time varying inductors on the pulse propagations over $50\;\mu s$ along the circuit chain for three settings of the compensation. The serial resistance is compensated down to (a) $1\;\Omega$. (b) $10\;\Omega$. (c) $50\;\Omega$. Depicted are the voltage signals at nodes one to eight counted from the node of signal input.}
    \label{fig:comp_Ohm}
\end{figure}

\begin{figure}
    \centering
    \includegraphics[width=0.7\linewidth]{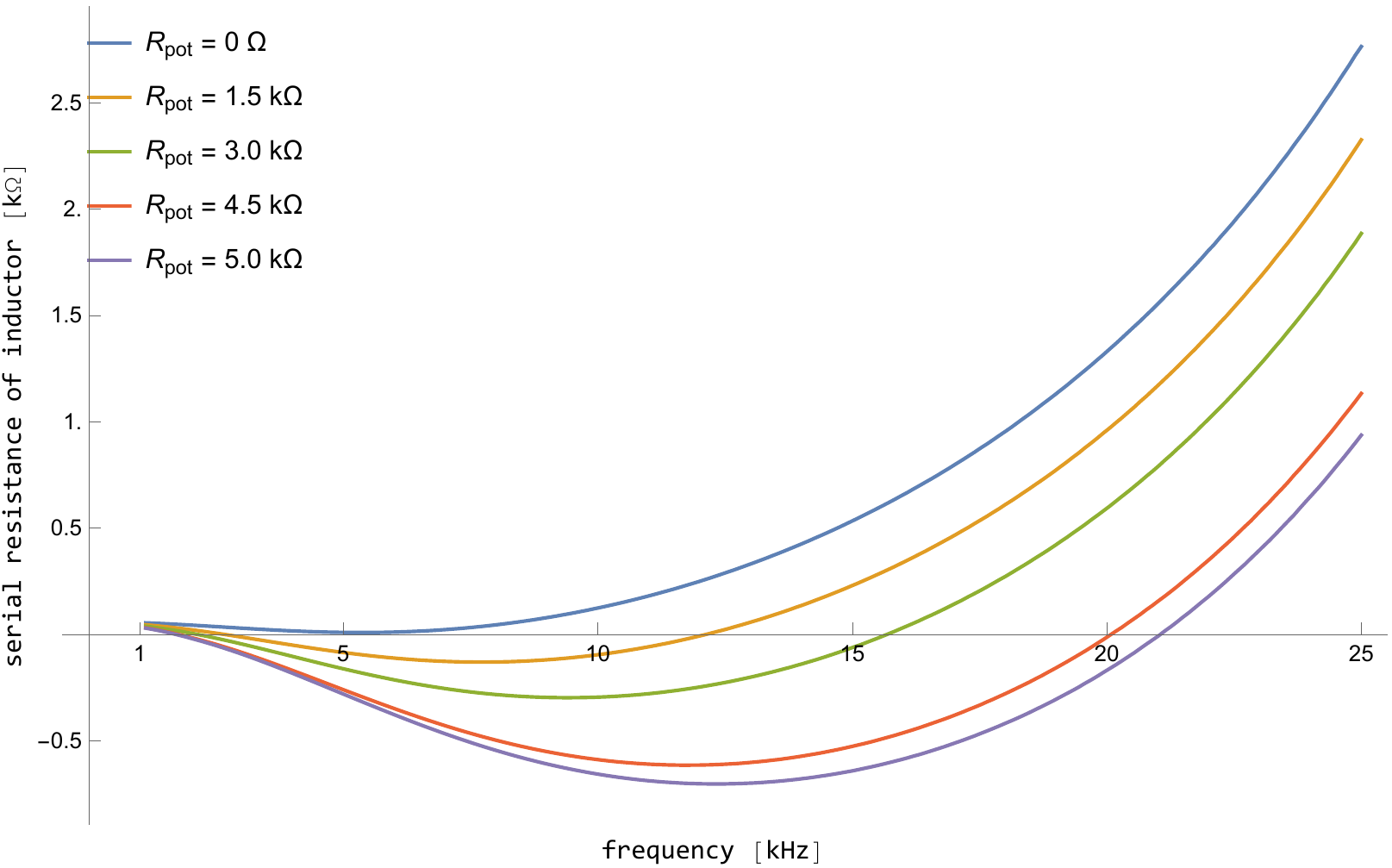}
    \caption{Influence of tuning the potentiometer's resistance of the compensation subcircuit (cf. Fig.\ref{fig:img9367} (10)) on the serial resistance of the inductor of a single node. For a resistance setting of the potentiometer of slightly above $0\;\Omega$ the resistance of the inductor gets very close to $0\;\Omega$ as well, allowing for the injected pulse to survive longer on it's way along the circuit chain. Negative resistance values, i.e. overcompensation, at the frequency of the injected pulse result in a gain of the signal and therefore in saturation of the multiplier output, which had to be avoided.}
    \label{fig:res_comp}
\end{figure}

The OpAmp acts as an analog integrator of the input voltage, its output voltage is given by $-\frac{1}{\mathrm{i}\omega\,R_I\,C_I} V_{\text{in}}.$ This signal is fed into the Z and X2 inputs of the analog multipliers, so that it is subtracted from the X1 input before multiplication and added to the output after multiplication. This way, the output W of the analog multiplier is 
\begin{align}
    W=&(V_\text{in} + \frac{1}{\mathrm{i}\omega\,R_I\,C_I} V_\text{in})f(\tau) - \frac{1}{\mathrm{i}\omega\,R_I\,C_I}V_\text{in}\\
    = &V_\text{in} [(1+\frac{1}{\mathrm{i}\omega\,R_I\,C_I})f(\tau) - \frac{1}{\mathrm{i}\omega\,R_I\,C_I}].
\end{align}
Finally, the output current flowing from the connected node through inductor $L$ with parasitic serial resistance $R_L$ is 
\begin{align}
I=&\frac{V_\text{in}-W}{\mathrm{i}\omega\,L + R_L}\\
= &\frac{1}{\mathrm{i}\omega L} \frac{(1+\frac{1}{\mathrm{i}\omega\,R_I\,C_I})(1-f(\tau))}{1+\frac{R_L}{\mathrm{i}\omega\,L}}V_\text{in}.
\end{align}
Now $R_I$ and $C_I$ are chosen such that $\frac{1}{R_I C_I} = \frac{R_L}{L}$ and we obtain
\begin{align}
    I=\frac{1}{\mathrm{i}\omega L} (1-f(\tau)) V_\text{in}.
\end{align}
This means that the added subcircuit can precisely compensate for the serial resistance of the inductor, which we consider to be the main cause of parasitic loss.
The influence of the serial resistance of the time varying inductor on the amplitude of the fed in pulse is shown in Fig.\,\ref{fig:res_comp}.
In our setup, an additional voltage divider consisting of two $\SI{100}{\kilo\ohm}$ resistors and a $\SI{5}{\kilo\ohm}$ potentiometer was included to fine-tune the output voltage of the integrator subcircuit.
Figure \,\ref{fig:comp_Ohm} shows how the different values of compensation affect the attenuation of a pulse pumped along the chain.
The influence of this compensation on the effective serial resistance of the inductor is shown in Fig.\,\ref{fig:res_comp} for different compensations, i.e. different resistance settings in the potentiometer, over frequency. 
For a resistance setting of the potentiometer of slightly above $0\;\Omega$ the resistance of the inductor gets very close to $0\;\Omega$ as well, allowing for the injected pulse to survive longer on it's way along the circuit chain.

\subsection{Measurement setup and procedure}\label{app:meas}

\begin{figure}[p]
	\centering
	\includegraphics[width=0.6\linewidth]{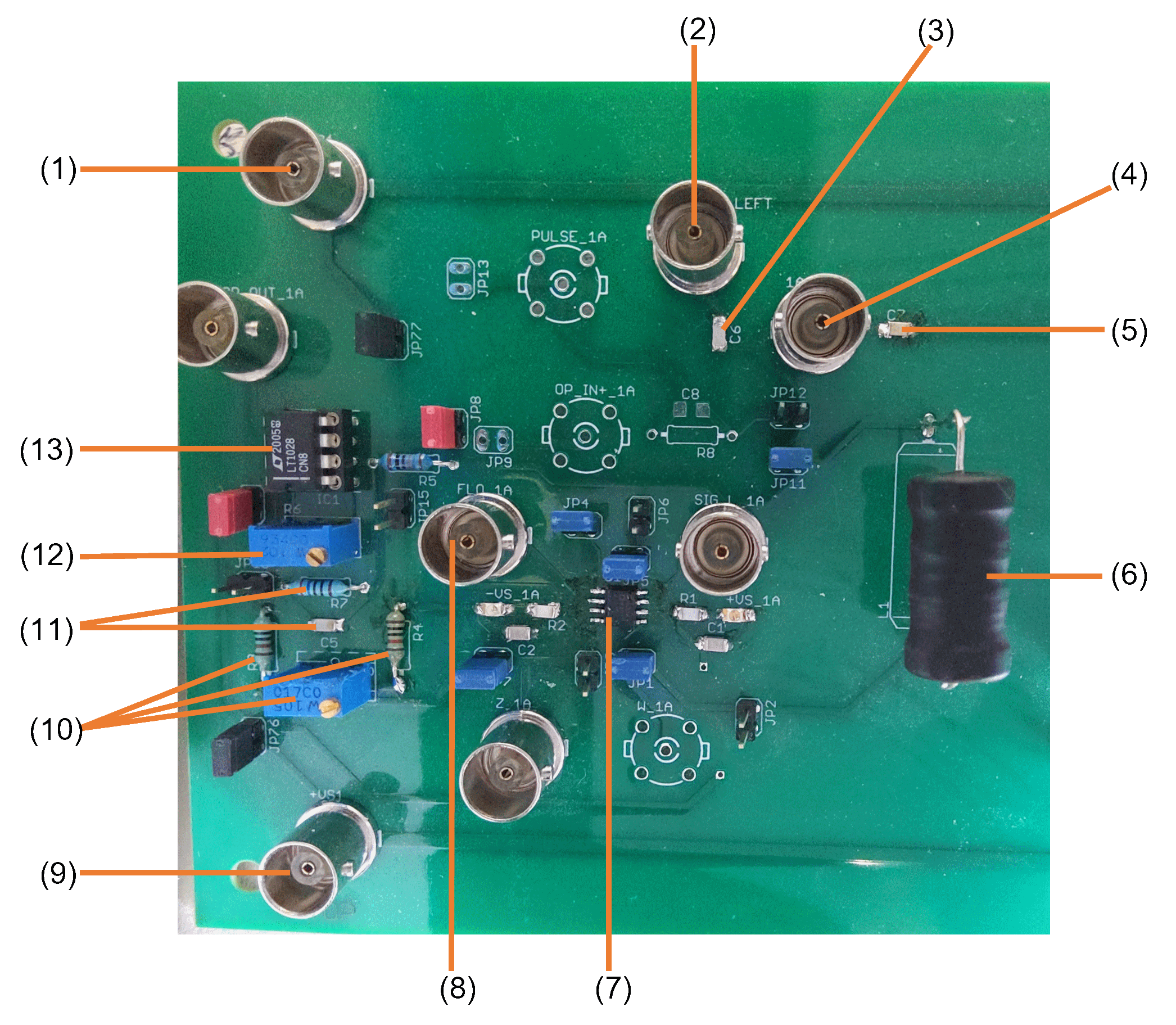}
	\caption{
 Single node of the Thouless pump circuit. (1) BNC connector for negative supply voltage. (2) BNC connector to previous board. (3) Hopping capacitor to previous node. (4) BNC connector to measured node. (5) Hopping capacitor to next node. (6) Inductor of measured node. (7) Analog multiplier providing time dependence for the inductor. (8) BNC connector to input Floquet signal into the multiplier. (9) BNC connector for positive supply voltage. (10) Resistors for gain adjustment to minimize serial resistance of the inductor. (11) Resistor and capacitor of the integrator built from the operational amplifier. (12) Potentiometer for offset adjustment of operational amplifier. (13) Operational amplifier used as part of an integrator in combination with the elements of (11).}
	\label{fig:img9367}
\end{figure}

For our experimental setup we devised printed circuit boards (PCBs).
Our chosen circuit elements are surface mounted capacitors for hopping between nodes, through hole inductors to ground, through hole resistors for tuning reasons (cf. Fig. \ref{fig:img9367} (10) and (11)) as well as one analog multiplier and one operational amplifier per node.
The multiplier is used to generate a time dependence in the behavior of the inductor, whereas the operational amplifier is part of an integrator circuit which is used to emulate nearly vanishing serial resistance of the inductor. 
This integrator was necessary to keep the measured pulses alive long enough to make several turns in the periodic chain.
To preserve translational symmetry the scatter of the absolute values of the circuit elements needed to be smaller than typical tolerances of commercially available components.
To this end all components were precharacterized by a BK Precision 894 LCR-meter.
The following choices of components were made for this experimental setup:
Inductor to ground: Bourns 5900-104-RC nominal values $ L = 100\; \text{mH}$ and $R_{\text{DC}} = 82\; \Omega$.
Hopping capacitor: Yageo CC0603GR-NPO-8BN102 nominal values $ C = 1\; \text{nF}$.
Integrator capacitor: Murata GJM1555-C1H470-GB01D nominal values $ C = 47\; \text{pF}$.
Resistor before integrator: Yageo MF0204 $R = 100\; \text{M}\Omega$.
Resistor within integrator: Yageo MF0204 $R = 100\; \text{M}\Omega$.
Resistors within voltage divider: Yageo MF0207 $R = 100\; \text{k}\Omega$.
Potentiometer within voltage divider: Bourns PV36-W502-C01B00 $R = 5\; \text{k}\Omega$.
The circuit was excited at the first node with a sinusoidal signal from an Agilent 33220A arbitrary wave form generator coupled inductively to the inductor of the first node using an inductor of the same kind as used on the  board.
The propagating wave fronts where measured with oscilloscopes of the PicoScope 4000 Series by PICOTech, connected to each node.
The impedance measurements at each node for the eigenfrequency band structures where performed with an MFIA by Zurich instruments.
For this measurement the Floquet phases where chosen stationary with a difference of $2\pi / 5$ from one node to the next.
The whole $2\pi$ of the Floquet phase was then sampled in $100$ equidistant points, leading to the figures shown in the main text.